\documentclass[10.5pt,a4paper,oneside]{extarticle}
\usepackage[T1]{fontenc} 
\usepackage[left=25mm,right=20.9mm,top=4.25cm,bottom=3.75cm,headsep=0.25cm,footskip=0.25cm]{geometry}
\usepackage[english]{babel}
    \usepackage[affil-it]{authblk}
    \usepackage{etoolbox}
    \usepackage{lmodern}
    \usepackage{setspace}
    \usepackage{amssymb,amsthm,amsmath}
    \usepackage{tikz}
    \usepackage{listings}
    \usepackage{color}
    \usepackage{caption}
    \usepackage{placeins}
    \usepackage{booktabs}
    \usepackage{enumerate}
    \usepackage{graphicx}
    \usepackage{titlesec}
    \usepackage{mathptmx}
    \usepackage{anyfontsize}
    \usepackage{t1enc}
    \usepackage[utf8]{inputenc}
    \usepackage{fancyhdr}
    \usepackage{float,xspace}
    \usepackage{scalefnt}
	
\usepackage{soul} 
\usepackage{tabulary}
\usepackage[colorlinks=true,linkcolor=blue]{hyperref} 

\usepackage{textcase}

\usepackage[colorinlistoftodos]{todonotes}

\titlespacing*{\section}
{0pt}{12pt}{0pt}
\titlespacing*{\subsection}
{0pt}{12pt}{0pt}
\titlespacing*{\subsubsection}
{0pt}{12pt}{0pt}
\titleformat{\section}
  {\huge\sffamily\Large\bfseries\color{black}}
  {\thesection}{1em}{}
    \titleformat{\subsection}[block]{\bfseries}{\thesubsection}{.5em}{}
    \titleformat{\subsubsection}[block]{\bfseries}{\thesubsubsection}{.5em}{}

\titleformat{\section}{\fontsize{12}{19}\bfseries}{\thesection}{1em}{}

\usepackage{mathptmx} 

\makeatletter

\patchcmd{\@maketitle}{\LARGE \@title}{\fontsize{14}{19.2}\selectfont\@title}{}{} 
\makeatother
\usepackage{siunitx}
\usepackage{subcaption}
\usepackage{bm}
\usepackage{bbm}
\newcommand{\e}{\bm{e}}
\renewcommand{\d}{\bm{d}}
\renewcommand{\r}{\bm{r}}
\newcommand{\x}{\bm{x}}
\newcommand{\y}{\bm{y}}
\newcommand{\G}{\bm{G}}
\newcommand{\W}{\bm{W}}
\newcommand{\A}{\bm{A}}

\newcommand{\Z}{\bm{Z}}
\renewcommand{\j}{\mathrm{j}}

\renewcommand{\H}{\mathsf{H}}
\newcommand{\Trans}{\mathsf{T}}
\newcommand{\Aint}{\A_{\mathrm{int}}}
\newcommand{\Aext}{\A_{\mathrm{ext}}}
\newcommand{\Jint}{\mathcal{J}_{\mathrm{int}}}
\newcommand{\Jext}{\mathcal{J}_{\mathrm{ext}}}
\newcommand{\Jpenal}{\mathcal{J}_{\mathrm{penal}}}
\newcommand{\Jconst}{\mathcal{J}_{\mathrm{const}}}
\newcommand{\Pred}{P_{\mathrm{red}}}
\title
{
	\vspace{-5cm}
	\begin{minipage}{\textwidth}	
	\hspace{-20pt}\vspace{100pt}
	\hspace{-65pt}
	\end{minipage}
	\textbf{Kernel-interpolation-based spatial active noise control \\ with exterior radiation suppression}
	\author[ ]{Kazuyuki ARIKAWA, Shoichi KOYAMA, and Hiroshi SARUWATARI$^{(1)}$}
  	\affil[(1)]{Graduate School of Information Science and Technology, The University of Tokyo, Japan, ak1217@g.ecc.u-tokyo.ac.jp}
}
\date{}

\begin{document}

\clearpage
\setcounter{page}{1}
\maketitle
\thispagestyle{empty}
\fancypagestyle{empty}
{	
	\fancyhf{} \fancyfoot[R]
	{
		\vspace{-2cm}
	}
}
\subsection*{\fontsize{10.5}{19.2}\uppercase{\textbf{Abstract}}}
{\fontsize{10.5}{60}\selectfont
A spatial active noise control (ANC) method based on kernel interpolation of a sound field with exterior radiation suppression is proposed. The aim of spatial ANC is to reduce incoming noise over a target region by using multiple secondary sources and microphones. The method based on kernel interpolation of a sound field allows noise attenuation in a regional space with an array of arbitrary geometry. The cost function is defined as the acoustic potential energy, i.e., the regional integral of the power distribution inside the target region. However, this cost function does not take into consideration the exterior radiation of secondary sources. Thus, the acoustic power in the exterior region can be amplified by the output of the secondary sources. We propose two spatial ANC methods with exterior radiation suppression. The first approach is based on the minimization of the cost function formulated as a sum of the interior acoustic potential energy and exterior radiation power. The second approach is based on the minimization of the interior acoustic potential energy with inequality constraints on the exterior radiation power.
Adaptive algorithms for minimizing the cost function are derived for the two approaches. Numerical experimental results indicate that the proposed methods can reduce the interior regional noise while suppressing the exterior radiation.}

\noindent{\\ \fontsize{11}{60}\selectfont Keywords: Active noise control, Kernel interpolation, Exterior radiation, Adaptive filtering}

\fontdimen2\font=4pt
\section{\uppercase{Introduction}}
Active noise control (ANC) is aimed at canceling incoming primary noise by emitting anti noise signals from secondary sources~\cite{kajikawa2012recent, kuo1999active, Nelson:ACS}. 
Multichannel ANC algorithms are typically applied to attenuate the primary noise at multiple control points by minimizing the power of the signals observed at error microphones. However, this conventional ANC takes only discrete positions into consideration; therefore, there is no guarantee that the primary noise is reduced in the region between the positions. 

Sound field recording and reproduction techniques~\cite{Koyama:IEEE_J_ASLP2013, poletti2005three,spors2008theory, Ueno:IEEE_SPL2018,Ueno:IEEE_J_SP_2021} have led to the recent development of spatial ANC methods aimed at attenuating noise in a continuous target region~\cite{Bu:ACM2018, Ito:ICASSP2019, Koyama:IEEE_ACM_J_ASLP2021, maeno2019spherical, Sun:ICASSP2019, zhang2016multichannel}.
In particular, spatial ANC methods based on the kernel interpolation of the sound field~\cite{Ito:ICASSP2019, Ito:ICASSP2020} have the advantages that arbitrary configuration is allowed for the microphone array placement and there is no need for the prior identification stage required in virtual sensing techniques~\cite{Moreau:2008}.

Meanwhile, if we consider applications such as noise reduction inside automobile cabins~\cite{samarasinghe2016recent}, it is desirable to suppress the exterior radiation from the secondary sources to a certain level so that the secondary sources do not produce large noise outside the target region. 
The cost function used in the previous kernel-interpolation-based ANC method is the acoustic potential energy inside the target region; therefore, the previous method does not take into consideration the exterior radiation.  
Although ANC methods with a constraint on the output power of secondary sources have been investigated~\cite{qiu2001study, rafaely2000computationally, shi2019two, shi2021optimal}, restricting the output power of secondary sources does not necessarily lead to the suppression of exterior radiation in the spatial ANC.
We propose ANC methods to suppress exterior radiation from secondary sources while reducing noise inside the target region. We consider two approaches: the first approach is to add a penalty term corresponding to exterior radiation power to the cost function used in the previous study, i.e., the acoustic potential energy inside the target region, and the second approach is to put inequality constraints on the external radiation power and minimize the same cost function as in the previous method. We formulate normalized least-mean-square (NLMS) algorithms in the frequency domain to adaptively update the control filter for both approaches. Numerical experiments show that the proposed methods can reduce noise inside the target region while maintaining less exterior radiation than in the previous kernel-interpolation-based method. 

\section{\uppercase{Spatial Active Noise Control Based on Kernel Interpolation}}
\begin{figure}[tb]
  \begin{center}
    \includegraphics[width=0.5\columnwidth]{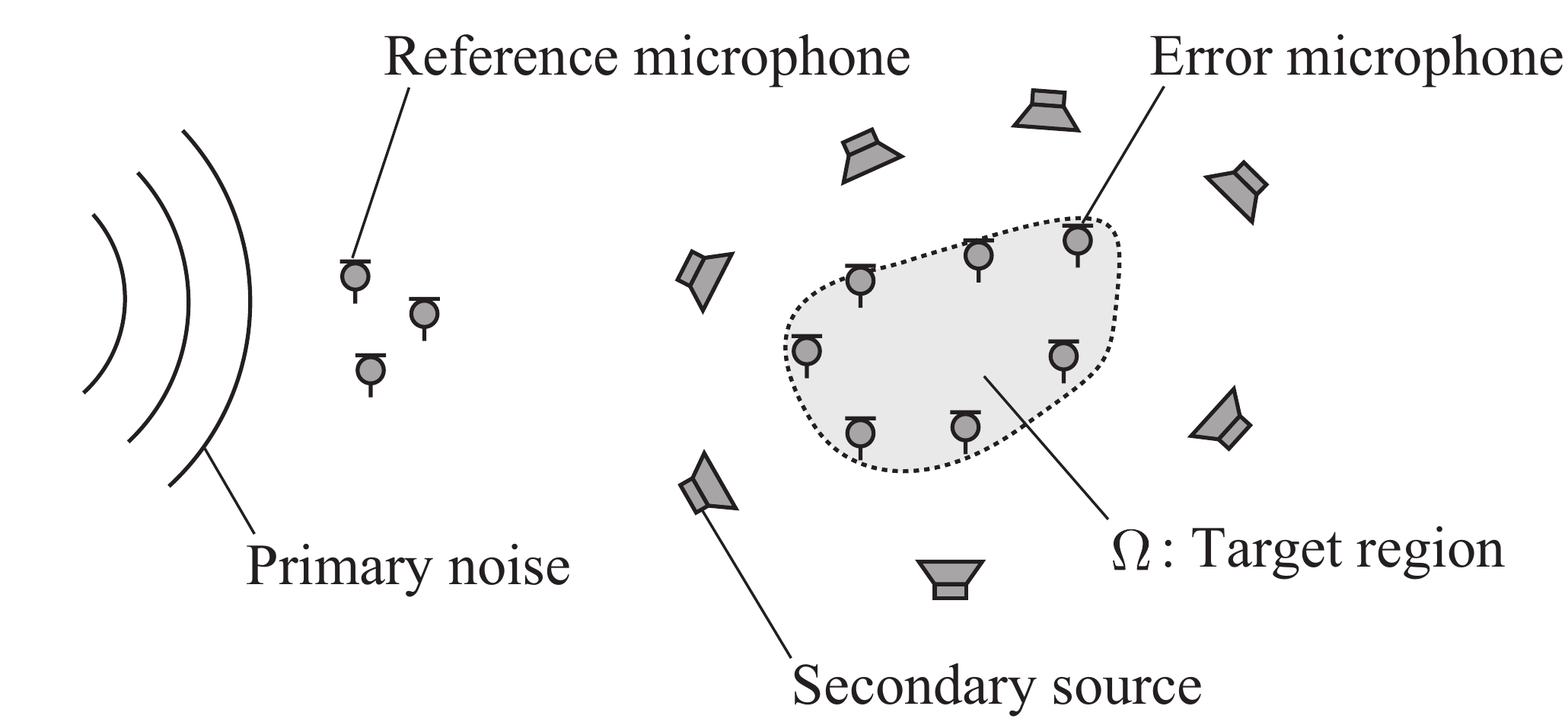}
  \end{center}
  \caption{Arrangement of spatial ANC system.}
  \label{fig:01}
\end{figure}

Let $\Omega \subset D$ be a target region where $D = \mathbb{R}^2$ or $\mathbb{R}^3$ and $M$ error microphones are placed in $\Omega$. $L$ secondary sources and $R$ reference microphones are placed outside $\Omega$, as shown in Fig.~\ref{fig:01}. 

The driving signals of the secondary sources and the observed signals of the reference and error microphones at angular frequency $\omega$ are denoted as $\y(\omega) \in \mathbb{C}^L$, $\x(\omega) \in \mathbb{C}^R$, and $\e(\omega) \in \mathbb{C}^M$, respectively. Hereafter, we omit the argument $\omega$ for notational simplicity. The goal of spatial ANC is to reduce primary noise in $\Omega$ by appropriately setting the driving signals $\y$ based on the observed signals $\x$ and $\e$. When denoting the primary noise at the error microphone positions as $\d \in \mathbb{C}^M$, the error microphone signals are represented as
\begin{align}
  \e = \d + \G \y,
\end{align}
where $\G \in \mathbb{C}^{M\times L}$ consists of the transfer functions from the secondary sources to the error microphones, which we assume to be given in advance. The driving signals are calculated as
\begin{align}
  \y = \W \x, \label{eq:Wx}
\end{align}
where $\W \in \mathbb{C}^{L \times R}$ is the control filter matrix, which is adaptively updated to minimize a certain cost function $\mathcal{J}$.

In spatial ANC based on kernel interpolation~\cite{Ito:ICASSP2019}, the cost function is defined as the acoustic potential energy in the interior region of $\Omega$ as
\begin{align}
    \Jint := \int_{\Omega}|\hat{u}_{\mathrm{e}}(\bm{r})|^2 \,d\bm{r}, \label{eq:cost}
\end{align}
where $\hat{u}_{\mathrm{e}}(\r)$ is the pressure field inside $\Omega$ estimated by kernel ridge regression. By using the estimation formula of $\hat{u}_{\mathrm{e}}(\r)$ described in~\cite{Ito:ICASSP2019}, $\Jint$ can be represented as
\begin{align}
    \Jint &= \e^{\H} \Aint \e, \label{eq:cost_prev}
\end{align}
where $\Aint$ is the Hermitian matrix defined as
\begin{align}
    \Aint := \bm{P}^{\H}\left(\int_{\Omega}\bm{\kappa}^{\ast}(\r)\bm{\kappa}^{\Trans}(\r)\,\mathrm{d}\r \right)\bm{P}. \label{eq:Aint}
\end{align}
Here, $(\cdot)^{\H}$ denotes the complex conjugate and $\bm{P} = (\bm{K} + \lambda \bm{I}_M)^{-1}$, where $\bm{K} \in \mathbb{C}^{M \times M}$ and $\bm{\kappa}(\r) \in \mathbb{C}^M$ are the Gram matrix and the vector consisting of the kernel function $\kappa(\cdot, \cdot)$, respectively. The kernel function is defined as
\begin{align}
    \kappa(\r, \r^{\prime}) := \begin{cases} 
    J_0(k\|\r - \r^{\prime}\|_2) & \text{if $D = \mathbb{R}^2$} \\
    j_0(k\|\r - \r^{\prime}\|_2) & \text{if $D = \mathbb{R}^3$}
    \end{cases}, \label{eq:kappa}
\end{align}
where $k$ denotes the wave number, $\|\cdot\|_2$ denotes the $\ell_2$-norm, and $J_0(\cdot)$ and $j_0(\cdot)$ are the 0th-order Bessel and spherical Bessel functions of the first kind, respectively. The NLMS algorithm for updating $\W$ to minimize $\Jint$ is derived as
\begin{align}
    \W_{n+1} &= \W_n - \mu_n \left.\frac{\partial}{\partial \W^{\ast}}\right|_{\W = \W_n}\Jint \notag \\
    &= \W_n - \frac{\mu_0}{\|\G^{\H}\Aint\G\|_2\|\x_n\|_2^2 + \epsilon}\G^{\H}\Aint\e_n\x_n^{\H}, \label{eq:prev_anc}
\end{align}
where $\mu_0 \in (0, 2)$ is the normalized step-size parameter, and $\epsilon > 0$ is a regularization parameter to avoid zero division. $\x_n$ and $\e_n$ are the observed signals at the $n$th iteration. 

\section{\uppercase{Spatial Active Noise Control with exterior radiation suppression}}
Noise reduction inside the target region is achieved by minimizing $\Jint$ in \eqref{eq:cost_prev}, but it can possibly lead to unacceptably large exterior radiation power from the secondary sources. 
Now, we formulate two adaptive filtering algorithms to suppress exterior radiation power while reducing noise in the target region. 
The representation of exterior radiation power from the secondary sources is introduced in Section~\ref{subsec:3-1}, and the NLMS algorithms with a penalty term or inequality constraints are derived in Sections~\ref{subsec:3-2} and \ref{subsec:3-3}, respectively.

\subsection{Derivation of exterior acoustic radiation}\label{subsec:3-1}
Let $\Omega_{\mathrm{S}}$ be a circular or spherical area including $\Omega$ and all the secondary sources. The total acoustic power radiated by the secondary sources, or the exterior radiation power, is defined as~\cite{Ueno:ICASSP2018}
\begin{align}
    \Jext := \int_{\partial \Omega_{\mathrm{S}}}\frac{1}{2}\mathrm{Re}\left[u_{\mathrm{s}}(\r)\frac{\j}{\rho c k}\frac{\partial}{\partial \bm{n}}u_{\mathrm{s}}(\r)^{\ast}\right]\,\mathrm{d}\r, \label{eq:Jext}
\end{align}
where $\rho$ is the density of the air, $c$ is the sound speed, $\mathrm{Re}[\cdot]$ denotes the real part of the complex number, $u_{\mathrm{s}}(\r)$ is the pressure field produced by the secondary sources, and $\partial/\partial \bm{n}$ denotes the normal derivative. By representing $u_{\mathrm{s}}(\r)$ with the driving signals $\y$, $\Jext$ can be expressed as~\cite{Ueno:ICASSP2018}
\begin{align}
    \Jext = \y^{\H}\Aext \y. \label{eq:Jext_quad}
\end{align}
Here, $\Aext$ is the Hermitian matrix determined by the positions and directivity patterns of the secondary sources. When all the secondary sources are point sources, each element of $\Aext$ can be represented as
\begin{align}
    (\Aext)_{ll^{\prime}} = \begin{cases} 
    \displaystyle\frac{1}{8\rho c k}J_0(k\|\r_{l} - \r_{l^{\prime}}\|_2) & \text{if $D = \mathbb{R}^2$}\\
    \displaystyle\frac{1}{8\rho c k}j_0(k\|\r_{l} - \r_{l^{\prime}}\|_2) &\text{if $D = \mathbb{R}^3$}
    \end{cases}, \label{eq:Aext}
\end{align}
where $\r_l$ is the position of the $l$th secondary source. The general representation for arbitrary directivity patterns is also shown in \cite{Ueno:ICASSP2018}.

\subsection{NLMS algorithm with penalty term}\label{subsec:3-2}
To minimize sound pressure inside $\Omega$ with suppression of the external radiation from the secondary sources, we define a new cost function $\Jpenal$ as
\begin{align}
    \Jpenal := \Jint + \lambda \Jext, \label{eq:cost_nlms_ext}
\end{align} 
where $\lambda > 0$ is the regularization parameter for the penalty term in $\mathrm{kg} / \mathrm{s}$. By substituting $\Jpenal$ instead of $\Jint$ in the derivation of Eq.~\eqref{eq:prev_anc}, the NLMS algorithm with a penalty term for exterior radiation is derived as
\begin{align}
    \W_{n+1} &= \W_n - \mu_n (\G^{\H}\Aint\e_n + \lambda \Aext\y_n) \x_n^{\H}. \label{eq:nlms_ext}
\end{align}
Here, $\mu_n$ is the step size at the $n$th iteration and is defined as
\begin{align}
    \mu_n &= \frac{\mu_0}{\|\G^{\H}\Aint\G + \lambda \Aext\|_2\|\x_n\|_2^2 + \beta}, \label{eq:step_nlms_ext}
\end{align}
where $\mu_0$ and $\beta$ are the same parameters as in Eq.~\eqref{eq:prev_anc}.

\subsection{NLMS algorithm with inequality constraints}\label{subsec:3-3}
The difficulty with the NLMS algorithm with a penalty term is the necessity of an appropriate setting of the regularization parameter $\lambda$. If $\lambda$ is large, sound pressure inside $\Omega$ will not be sufficiently reduced. In contrast, if $\lambda$ is small, the external radiation power can be too large. Another difficulty is that the external radiation can become large during the adaptation process even if the external radiation after the convergence of the control filter is acceptable.

To overcome the above difficulties, we reformulate the spatial ANC problem as an optimization problem with inequality constraints as
\begin{align}
\begin{aligned}
& \underset{{\W}\in{\mathbb{C}^{L \times R}}} {\text{minimize}} &&\Jint(\W)\\
&\text{subject to} && \Jext(\W) \leq C.
\end{aligned}\label{eq:ineq}
\end{align}
We here explicitly represent $\Jint$ and $\Jext$ as functions of the control filter $\W$. $C > 0$ is the maximum acoustic power radiated by the secondary sources. The optimization problem in Eq.~\eqref{eq:ineq} is equivalent to the minimization problem of the cost function $\Jconst$, which is defined as
\begin{align}
    \Jconst(\W) := \Jint(\W) + \mathbbm{1}_{\{\W \in \mathbb{C}^{L \times R} \mid \Jext(\W) \leq C \}}(\W). \label{eq:Jprox}
\end{align}
Here, $\mathbbm{1}(\cdot)$ is the indicator function, which returns $0$ if $\Jext(\W) \leq C$ and $\infty$ otherwise. Both terms in the right side of Eq.~\eqref{eq:Jprox} are convex and $\Jint$ is differentiable with respect to $\W$; therefore, we can apply the proximal gradient method~\cite{combettes2005signal} to minimize $\Jconst$. The update rules, which we call the NLMS algorithm with inequality constraints for external radiation, are represented as
\begin{align}
    \Z_{n+1} &= \W_{n}
    - \mu \Aext^{-1}(\G^{\H}\Aint\e_n \x_n^{\H}) \left({\bm{R}_{xx}^{(n)}}\right)^{-1}\label{eq:prox_start} \\
    \Tilde{\y}_{n+1} &= \Z_{n+1}\x_n \\
    \W_{n+1} &= \min\left(1, \frac{\sqrt{C}}{\sqrt{\Tilde{\y}_{n+1}^{\H}\Aext\Tilde{\y}_{n+1}}}\right)\Z_{n+1}, \label{eq:prox_end}
\end{align}
where $\mu$ in Eq.~\eqref{eq:prox_start} is the step size determined as
\begin{align}
    \mu = \frac{\mu_0}{\|\Aext^{-1} \, \G^{\H}\Aint\G\|_2 + \beta}. \label{eq:step_prox}
\end{align}
Here, $\mu_0$ and $\beta$ are the parameters defined in the same way as in Eqs.~\eqref{eq:prev_anc} and \eqref{eq:step_nlms_ext}. 
${\bm{R}_{xx}^{(n)}}$ is the estimation of $\bm{R}_{xx}$ at the $n$th iteration, where $\bm{R}_{xx} := \mathbb{E}\left[\x \x^{\H}\right]$ is the autocorrelation matrix of the reference microphone signals. If the number of reference microphones $R = 1$, $\left({\bm{R}_{xx}^{(n)}}\right)^{-1}$ can be estimated as $1 / \|\x_n\|_2^2$. If $R \geq 2$, $\bm{\Lambda}_{xx}^{(n)} := \left({\bm{R}_{xx}^{(n)}}\right)^{-1}$ is estimated as
\begin{align}
    {{\bm{\Lambda}_{xx}^{(n)}}} &= (\alpha{\bm{R}_{xx}^{(n-1)}} + (1 - \alpha)\x_n \x_n^{\H})^{-1} \notag \\
    &= \frac{1}{\alpha} \left[{{\bm{\Lambda}_{xx}^{(n-1)}}} - \frac{({{\bm{\Lambda}_{xx}^{(n-1)}}}\x_n)({{\bm{\Lambda}_{xx}^{(n-1)}}}\x_n)^{\H}}{\x_n^{\H}{{{\bm{\Lambda}_{xx}^{(n-1)}}}}\x_n + \frac{\alpha}{1 - \alpha}}\right]. \label{eq:woodbury}
\end{align}
Here, $\alpha \in (0, 1)$ and we apply the Sherman-Morrison formula to derive Eq.~\eqref{eq:woodbury}. Furthermore, to avoid the instability of calculating the inverse of $\Aext$, we replace $\Aext$ in Eqs.~\eqref{eq:prox_start} and \eqref{eq:step_prox} with $\Aext + \eta \bm{I}_L\,(\eta > 0)$ if the condition number of $\Aext$ in the $\ell_2$-norm is larger than a certain threshold. 
\section{\uppercase{Experiments}}
\subsection{Settings}\label{subsec:setting}
\begin{figure}[t]
  \begin{center}
    \includegraphics[width=0.5\columnwidth]{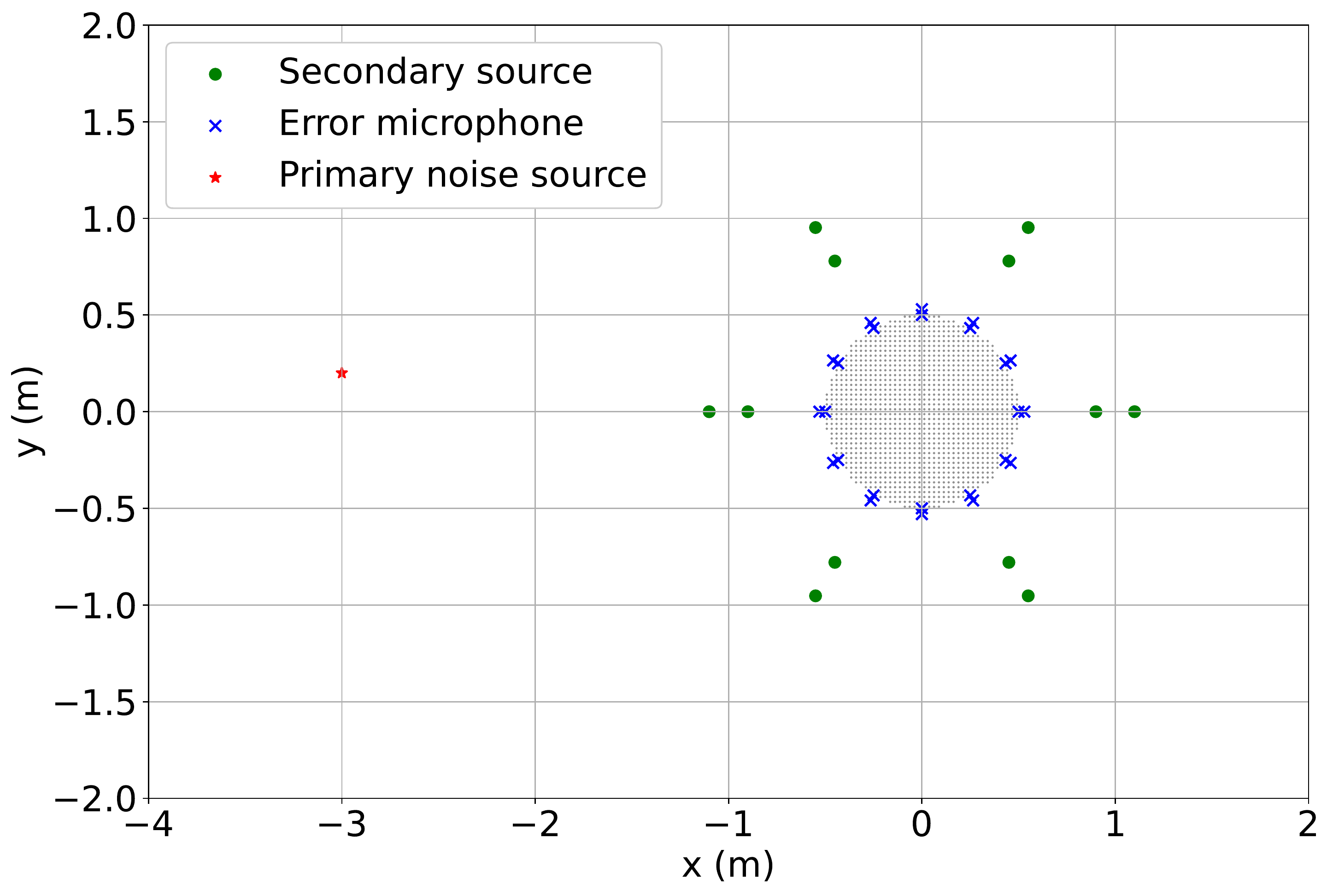}
  \end{center}
  \caption{Experimental setup. }
  \label{fig:schematic}
\end{figure}

We conducted numerical experiments to evaluate the proposed methods. We assumed a two-dimensional free field in the experiments. The target region $\Omega$ was a circular region with a radius of \SI{0.5}{m}, whose center was set at the coordinate origin, as shown in Fig.~\ref{fig:schematic}. The numbers of secondary sources and error microphones were set to $L=12$ and $M=24$, respectively. Secondary sources were placed at regular intervals on circles with radii of \SI{0.9}{m} and \SI{1.1}{m}, and error microphones were placed on circles with radii of \SI{0.47}{m} and \SI{0.53}{m}. A single source was placed as the primary noise source at (\SI{-3.0}{m}, \SI{0.2}{m}), and the reference microphone signal was directly obtained from the noise source. All secondary sources and the primary noise source were point sources. The noise signal is constant in the frequency domain, and Gaussian noise  was added to the error microphone and reference microphone signals at each iteration so that the signal-to-noise ratio became \SI{40}{dB}.

As an evaluation measure of ANC, we define the regional noise power reduction inside $\Omega$ as
\begin{align}
    \Pred(n) := 10\log_{10}\frac{\sum_j|u_{\mathrm{e}}^{(n)}(\r_j)|^2}{\sum_j|u_{\mathrm{p}}^{(n)}(\r_j)|^2},
\end{align}
where $\r_j$ is the $j$th evaluation point in $\Omega$, and $u_{\mathrm{e}}^{(n)}$ and $u_{\mathrm{p}}^{(n)}$ represent the total pressure field and primary noise field at the $n$th iteration, respectively. We set 1240 evaluation points at regular intervals inside $\Omega$. The external radiation power from secondary sources was calculated using Eq.~\eqref{eq:Jext_quad}.

We compared the NLMS algorithm with the penalty term for external radiation (\textbf{Ext-Penal NLMS}), the NLMS algorithm with inequality constraints for external radiation (\textbf{Ext-Const NLMS}), and the NLMS algorithm without exterior radiation suppression (\textbf{NLMS}). $\Aint$ was computed using Eq.~\eqref{eq:Aint} with numerical integration, and $\Aext$ was computed using Eq.~\eqref{eq:Aext}. The sound speed $c$ and the air density $\rho$ were set to \SI{340}{m/s} and \SI{1.3}{kg/m^3}, respectively. We replaced $\Aext$ with $\Aext + \eta \bm{I}_L$ in \textbf{Ext-Const NLMS} when the condition number of $\Aext$ exceeded $10^2$, where $\eta$ was set to $10^{-5}$. The parameters $\mu_0$ and $\beta$ in Eqs.~\eqref{eq:prev_anc}, \eqref{eq:step_nlms_ext}, and \eqref{eq:step_prox} were set to $0.9$ and $10^{-8}$, respectively. 

\begin{figure}[t]
     \centering
     \begin{subfigure}[t]{0.4\columnwidth}
         \centering
         \includegraphics[width=\columnwidth]{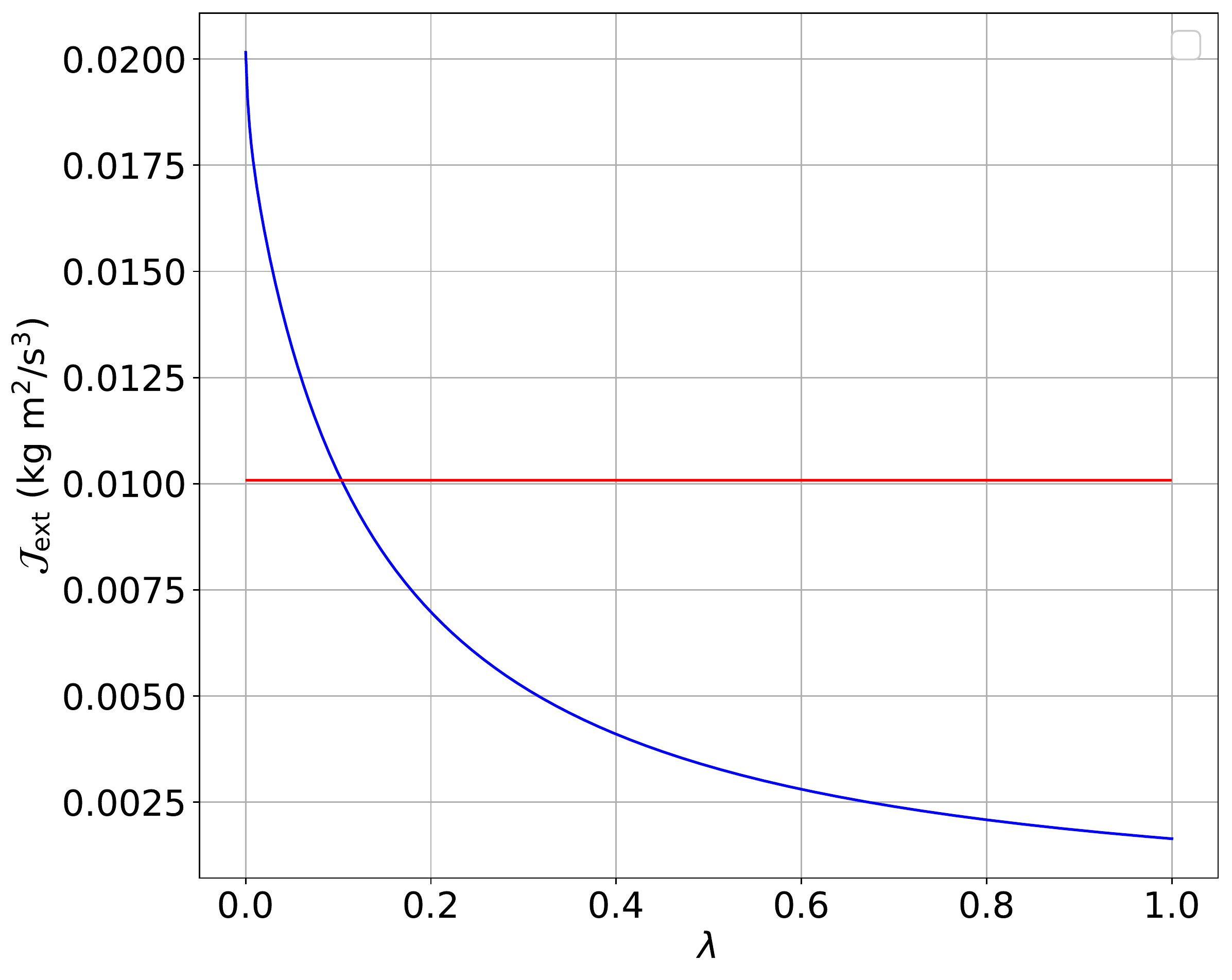}
         \caption{External radiation power. Red line indicates $(1/2)\hat{\mathcal{J}}_{\mathrm{ext}}$.}
         \label{fig:lambda_change_a}
     \end{subfigure}
     \hspace{6mm}
     \begin{subfigure}[t]{0.4\columnwidth}
         \centering
         \includegraphics[width=\columnwidth]{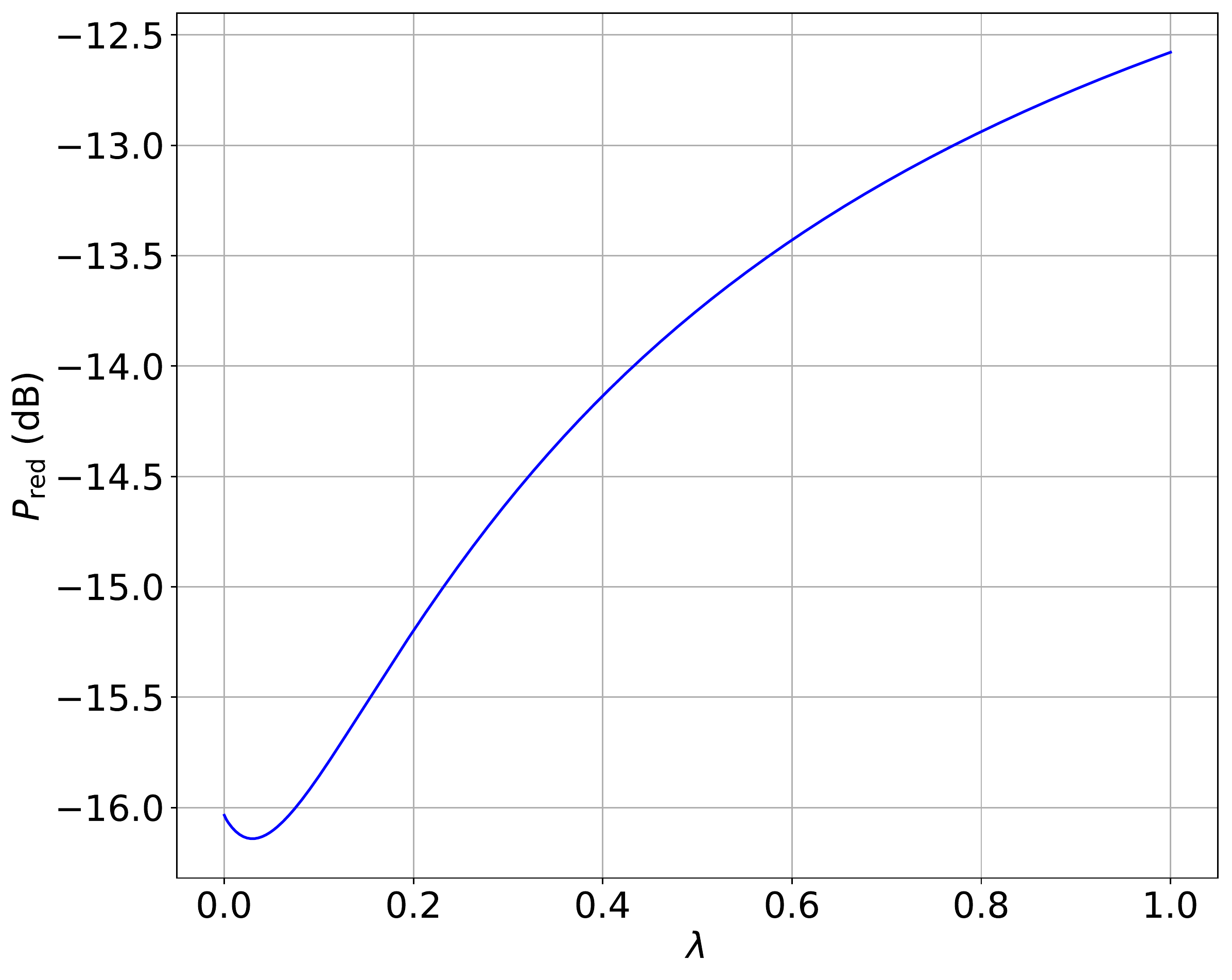}
         \caption{Regional noise power reduction}
         \label{fig:lambda_change_b}
     \end{subfigure}
     \caption{External radiation power and regional noise power reduction of \textbf{Ext-Penal NLMS} for each regularization parameter at \SI{600}{Hz}.}
     \label{fig:lambda_change}
\end{figure}

We denoted the external radiation power produced when the secondary sources were driven with the Wiener filter of \textbf{NLMS}, i.e., the convergence point of Eq.~\eqref{eq:prev_anc}, as $\hat{\mathcal{J}}_{\mathrm{ext}}$. The maximum radiation power $C$ used in \textbf{Ext-Const NLMS} was set to $(1/2)\hat{\mathcal{J}}_{\mathrm{ext}}$. We determined the regularization parameter $\lambda$ used in \textbf{Ext-Penal NLMS} by exhaustive search. Figures~\ref{fig:lambda_change_a} and \ref{fig:lambda_change_b} show the change of external radiation power and regional noise power reduction at \SI{600}{Hz} when $\lambda$ was varied from 0 to 1.0. Figure~\ref{fig:lambda_change_a} indicates that $\lambda$ should be set larger than \SI{0.1}{kg/s} to suppress the radiation power to less than $(1/2)\hat{\mathcal{J}}_{\mathrm{ext}}$. On the other hand, Fig.~\ref{fig:lambda_change_b} shows that the noise reduction performance monotonically decreases as $\lambda$ increases. In the experiments, we set $\lambda$ as $0.1$ at \SI{600}{Hz} so that the external radiation power after convergence became close to $(1/2)\hat{\mathcal{J}}_{\mathrm{ext}}$.
We note that the above procedure to select the parameter $\lambda$ is infeasible in practice since the Wiener filter cannot be computed in advance.

\subsection{Results}
\begin{figure}[t]
     \centering
     \begin{subfigure}[b]{0.49\columnwidth}
         \centering
         \includegraphics[width=\columnwidth]{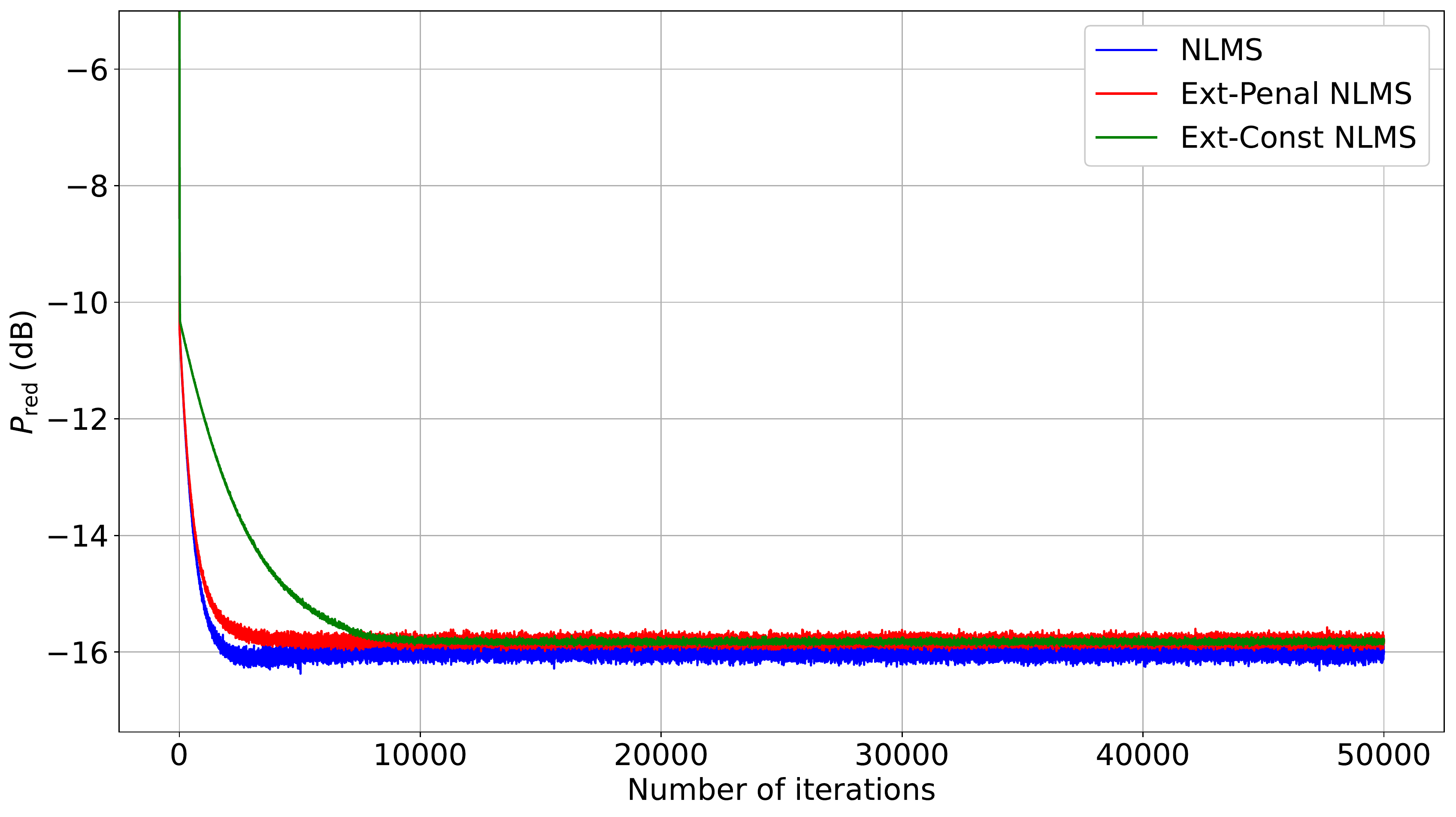}
         \caption{Regional noise power reduction}
         \label{fig:600Hz_result_a}
     \end{subfigure}
     \hfill
     \begin{subfigure}[b]{0.49\columnwidth}
         \centering
         \includegraphics[width=\columnwidth]{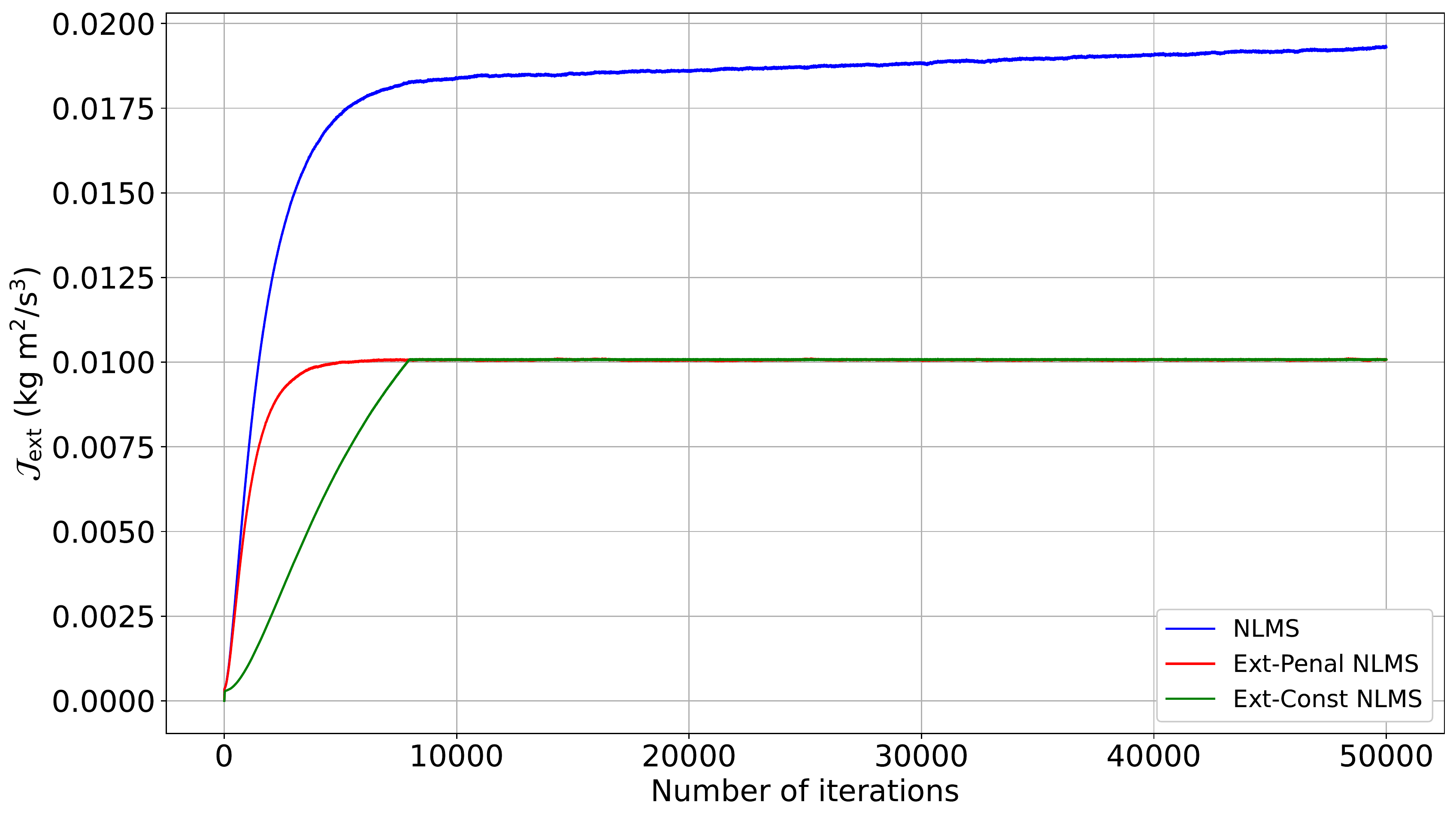}
         \caption{Exterior radiation power}
         \label{fig:600Hz_result_b}
     \end{subfigure}
     \caption{Regional noise power reduction and exterior radiation power at each iteration when noise frequency was \SI{600}{Hz}.}
     \label{fig:600Hz_result}
\end{figure}

\begin{figure}[t]
     \centering
     \begin{subfigure}[b]{0.45\columnwidth}
         \centering
         \includegraphics[width=\columnwidth]{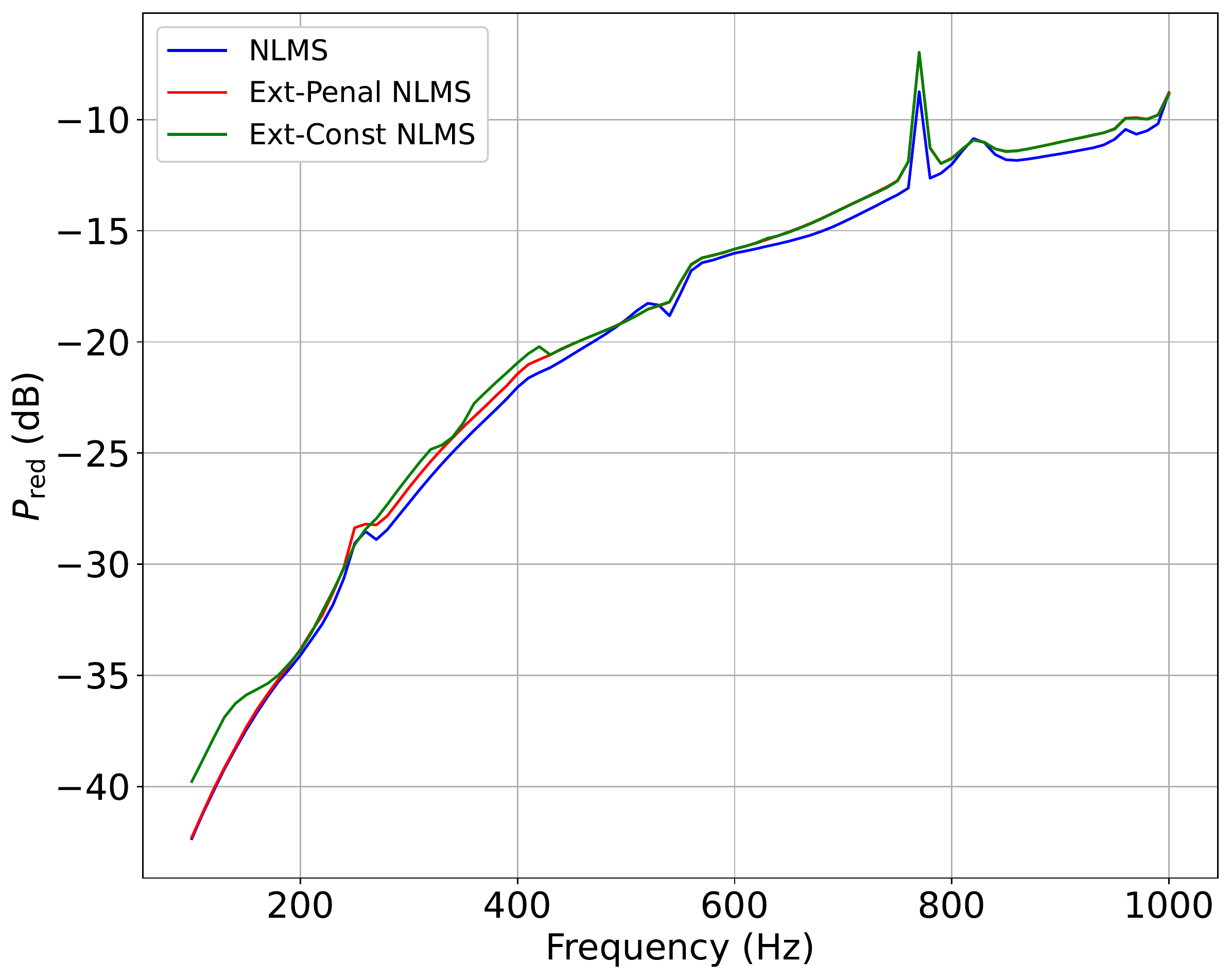}
         \caption{Regional noise power reduction}
         \label{fig:allHz_result_a}
     \end{subfigure}
     \hspace{10mm}
     \begin{subfigure}[b]{0.45\columnwidth}
         \centering
         \includegraphics[width=\columnwidth]{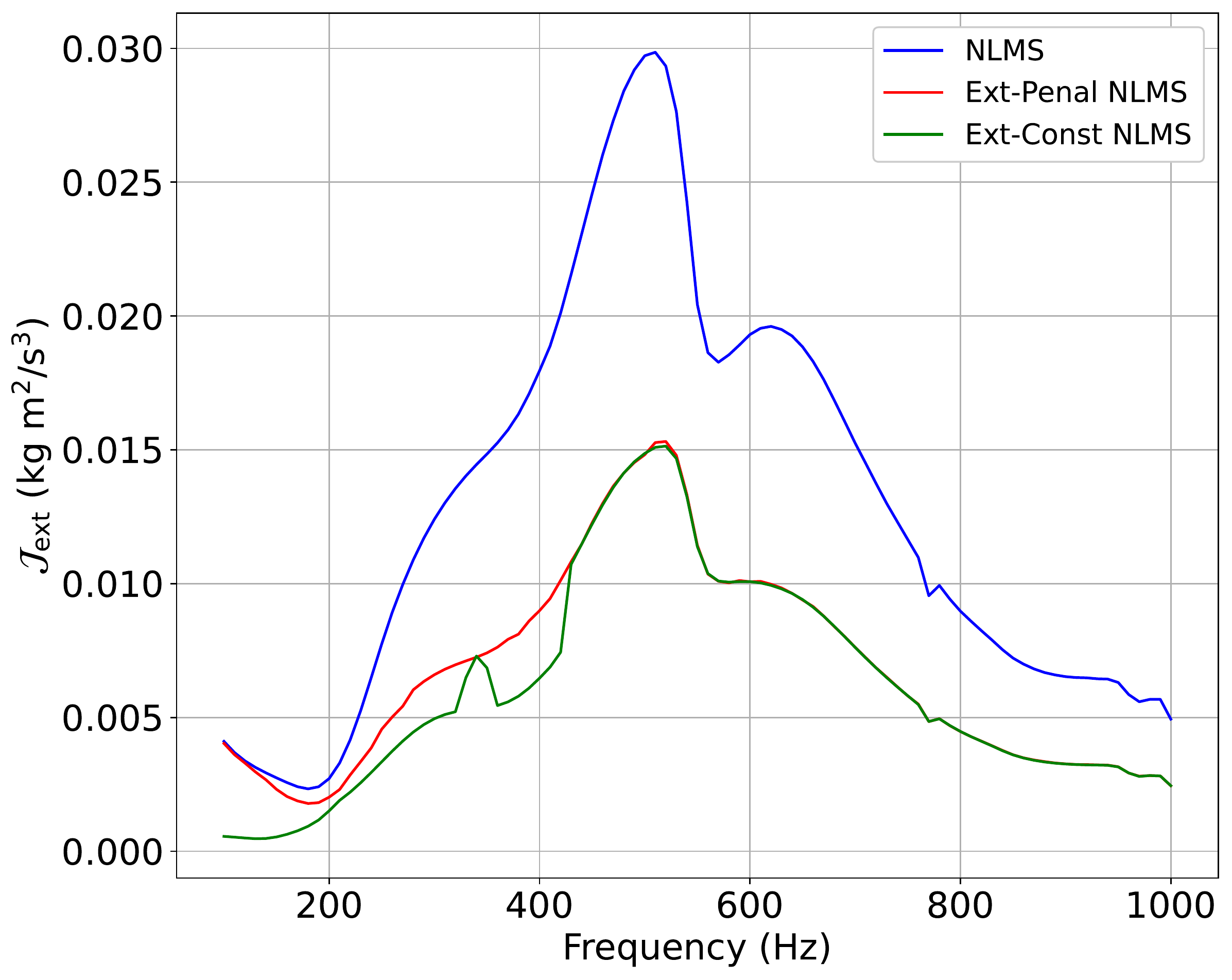}
         \caption{Exterior radiation power}
         \label{fig:allHz_result_b}
     \end{subfigure}
     \caption{Regional noise power reduction and exterior radiation power after 50000 iterations with respect to frequency.}
     \label{fig:allHz_result}
\end{figure}

Figure~\ref{fig:600Hz_result_a} shows the noise reduction performance at each iteration when the noise frequency was \SI{600}{Hz}. The final noise reduction performance was almost the same among the three methods, whereas the convergence of \textbf{Ext-Const NLMS} was slower than those of the other methods. This slow convergence is due to the smaller step size than in the other methods. Figure~\ref{fig:600Hz_result_b} shows the external radiation power at each iteration. The final radiation powers for \textbf{Ext-Penal NLMS} and \textbf{Ext-Const NLMS} were half of that for \textbf{NLMS}, as expected. 
In Fig.~\ref{fig:allHz_result_a}, $\Pred$ after 5000 iterations are plotted with respect to the frequency from 100 to \SI{1000}{Hz} at intervals of \SI{10}{Hz}. The parameter $\lambda$ for \textbf{Ext-Penal NLMS} was selected at each frequency in the same way as described in Section~\ref{subsec:setting}. It can be seen that the noise reduction performance was almost the same among the methods at most frequencies. Figure~\ref{fig:allHz_result_b} shows $\Jext$ after 50000 iterations. $\Jext$ for \textbf{Ext-Const NLMS} was smaller than that for \textbf{Ext-Penal NLMS} below \SI{420}{Hz}. It can be considered that the final output power of the secondary sources became smaller on replacing $\Aext$ with $\Aext + \eta \bm{I}_L$ in \textbf{Ext-Const NLMS} at frequencies below \SI{420}{Hz}.

\begin{figure}[t]
     \centering
     \begin{subfigure}[t]{0.49\columnwidth}
         \centering
         \includegraphics[width=\columnwidth]{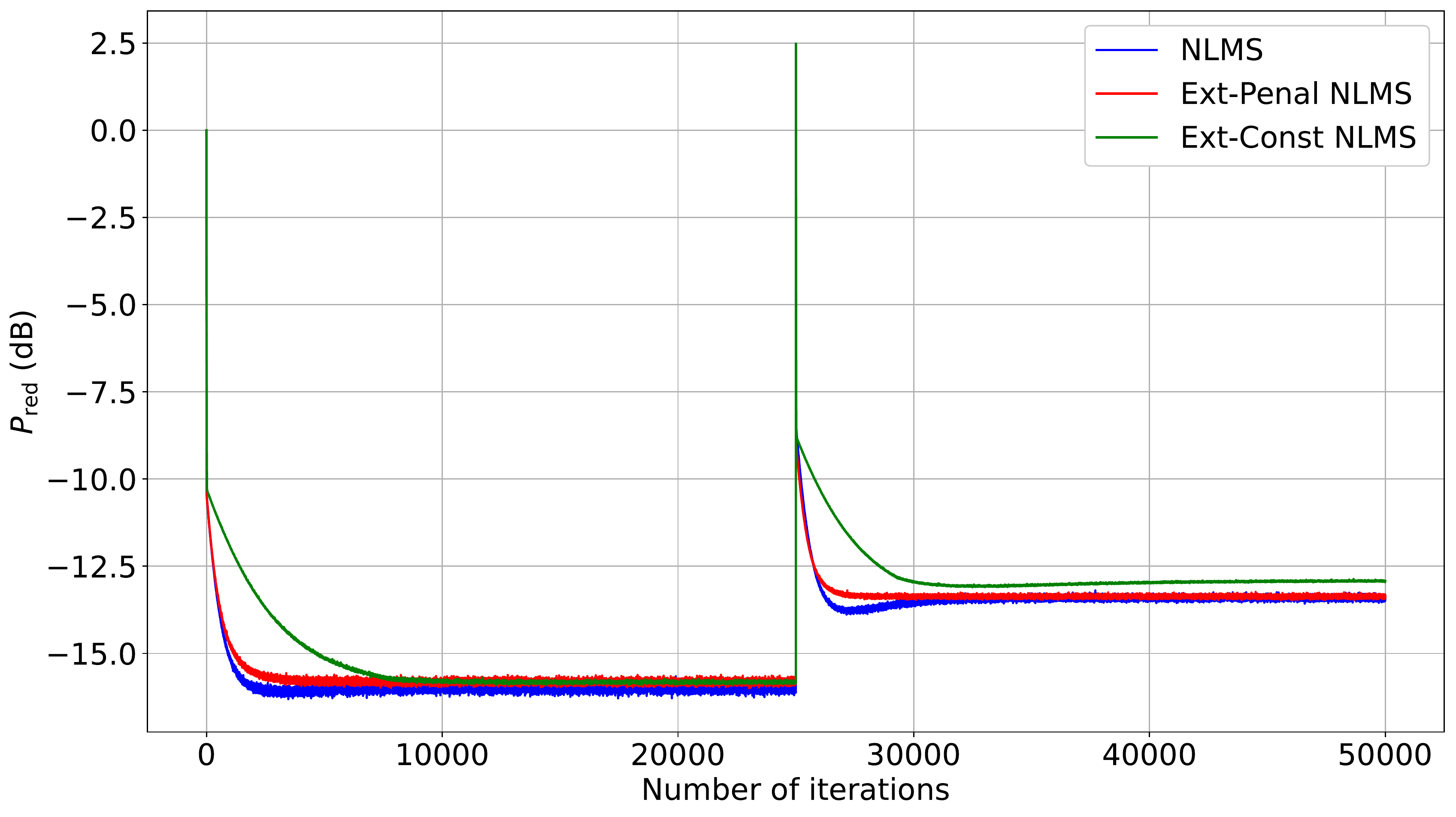}
         \caption{Regional noise power reduction}
         \label{fig:600Hz_move_result_a}
     \end{subfigure}
     \hfill
     \begin{subfigure}[t]{0.49\columnwidth}
         \centering
         \includegraphics[width=\columnwidth]{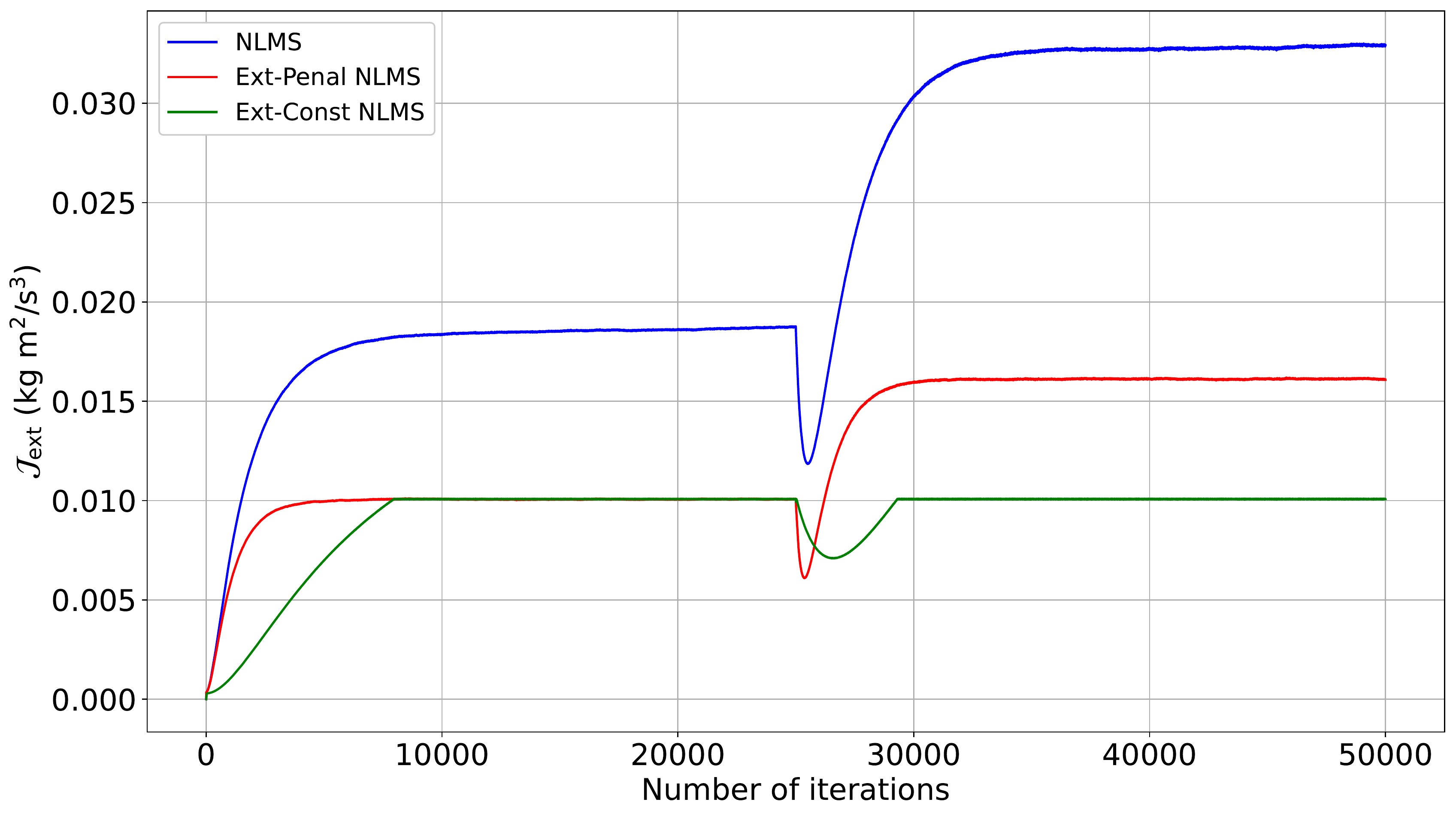}
         \caption{External radiation power}
         \label{fig:600Hz_move_result_b}
     \end{subfigure}
     \caption{Regional noise power reduction and exterior radiation power at each iteration when noise frequency was \SI{600}{Hz} and the position of the primary noise source changed at 25000 iterations.}
     \label{fig:600Hz_move_result}
\end{figure}

We also investigated the performance of the methods when the primary noise source moved during the adaptation process. The position of the primary noise source was (\SI{-3.0}{m}, \SI{0.2}{m}) for the first 25000 iterations, and (\SI{-2.0}{m}, \SI{0.2}{m}) for the next 25000 iterations. All the parameters were set the same as those described in Section~\ref{subsec:setting}.  Figure~\ref{fig:600Hz_move_result_a} shows $\Pred$ at each iteration when the frequency was \SI{600}{Hz}. Although $\Pred$ instantaneously increased when the position of the primary noise source changed at $n=25000$, the three methods successfully tracked the change of the noise source position. Figure~\ref{fig:600Hz_move_result_b} shows $\Jext$ at each iteration. For \textbf{NLMS} and \textbf{Ext-Penal NLMS}, $\Jext$ became larger after the primary noise source moved, whereas $\Jext$ remained the same for \textbf{Ext-Const NLMS}. The final exterior radiation powers for \textbf{NLMS} and \textbf{Ext-Penal NLMS} depend on the primary noise source signals, which are difficult to estimate. In contrast,
we can expect the same exterior radiation power for \textbf{Ext-Const NLMS} irrespective of the primary noise source signals, since the acceptable maximum external radiation power is determined in advance. 

\section{\uppercase{Conclusion}}
We proposed spatial ANC methods with the suppression of exterior radiation from secondary sources. We considered two cases: the addition of a penalty term corresponding to the exterior radiation power to the original cost function and the reformulation of the optimization problem with inequality constraints with respect to the external radiation power. The NLMS algorithms to update the control filter were derived for both cases. The NLMS algorithm with the penalty term converges faster than the NLMS algorithm with inequality constraints, whereas the latter method has the advantage that the maximum acceptable external radiation power can be set irrespective of the primary noise source signals.
The results of numerical experiments indicated that the proposed methods can reduce noise inside the target region while maintaining less exterior radiation than the previous method.

\section*{\uppercase{Acknowledgements}}
This work was supported by JST FOREST Program (Grant Number JPMJFR216M, Japan).
\bibliographystyle{abbrv}
\bibliography{str_def_abrv, reference, koyama_en_new}
\end{document}